\documentclass{article}

\usepackage{booktabs} %
\usepackage{balance}
\usepackage{enumitem}  %
\usepackage{subfigure}
\usepackage{xspace}
\usepackage{cleveref}
\usepackage{url}
\usepackage{graphicx}

\usepackage{fullpage}

\usepackage{makecell}

\usepackage[square,numbers]{natbib}
\newcommand{\etal}{\textit{et al}.\xspace}

\begin{document}

\title{A High School Camp on Algorithms and Coding in Jamaica}

\author{
  Daniel T. Fokum\thanks{The University of the West Indies, Mona.}\\
  \small{\url{daniel.fokum@uwimona.edu.jm}}
  \and
  Zaria Chen Shui\footnotemark[1]\\
  \small{\url{zaria.chenshui@mymona.uwi.edu}}
  \and
  Kerene Wright\footnotemark[1]\\
  \small{\url{kerene.wright@mymona.uwi.edu}}
  \and
  Orr Paradise\thanks{University of California, Berkeley.}\\
  \small{\url{orrp@eecs.berkeley.edu}}
  \and
  Gunjan Mansingh\footnotemark[1]\\
  \small{\url{gunjan.mansingh@uwimona.edu.jm}}
  \and
  Daniel Coore\footnotemark[1]\\
  \small{\url{daniel.coore@uwimona.edu.jm}}
}

\maketitle

\begin{abstract}
This is a report on JamCoders, a four-week long computer-science camp for high school
students in Jamaica. The camp teaches college-level
coding and algorithms, and targets academically excellent students in grades
9--11 (ages 14--17).

Qualitative assessment shows that the camp was, in general terms, a success. We
reflect on the background and academic structure of the camp and share key
takeaways on designing and operating a successful camp. We analyze data
collected before, during and after the camp and map the effects of
demographic differences on student performance in camp. We conclude with a
discussion on possible improvements on our approach.
\end{abstract}

\section{Introduction}
\label{sec:intro}

We report on \emph{JamCoders}, a four-week coding camp for high school students in Jamaica.
The camp targets a pool of
academically excellent, yet demographically diverse, students from
urban and rural areas across all parishes of Jamaica. There were no mathematical or programming prerequisites;
the only requirement on applicants was that they are in grade 9--11, i.e., ages 14--17.

The majority of the population of Jamaica
in which the camp was held are members of ethnic groups underrepresented in
Computing \cite{socsci7080122}. In fact, over 90\% of the population identifies as black. To increase Computing education opportunities
in Jamaica, it is important to understand the barriers affecting student
interest and retention.

Previous work shows that one potential factor contributing to the
under-representation of these groups may be limited early exposure to Computing
\cite{AlvaradoUM18, Parsons10, WangHRM16, FisherM03}.
Indeed, within Jamaica, most students are given the
option to graduate from high school with no Computing exposure. Another
known inhibiting factor is a high student-teacher ratio \cite{KOC201565,Meads0D23}.
The camp attempted to directly address these two factors by offering a summer
camp for high schoolers with a relatively low teacher-student ratio of 1:4.
The camp experience, provided at no cost to the students, included meals, lodging
and a Computing lab hosted on a local university campus.

The camp succeeded in securing applications from all parishes of Jamaica,
with 51\% of these applicants being female students. The admission process
ensured that there was gender balance and equitable representation of students
from across the country.

The camp required a significant investment of financial and human
resources. Our goal in this report is to communicate the structure and
implementation of this ambitious project, and analyse data collected
from the camp so as to understand how these resources could be better
invested in the future. We hope that this report informs other educators who
are considering organizing a similar camp in the future, as well as
scholars researching Computing education in similar settings to ours.

\paragraph{Outline}
\Cref{sec:litreview} reviews related work on  middle school and high school coding
camps. \Cref{sec:Design}
provides details about this camp, our research questions, and the collected data. \Cref{sec:rslts} offers statistical analysis of
this data, and \Cref{sec:Discuss} discusses our findings. In
\Cref{sec:limits}, we share limitations of our work. We provide
concluding remarks in \Cref{sec:concl}.

\section{Related Work}
\label{sec:litreview}

The use of summer camps to introduce coding and technology to high school
students has been explored by many Computer Science education researchers
\cite{Eckroth18, Fronza22}. Several coding camps were developed with the
intention of supporting minorities within Computing and were found to be
successful in targeting these groups through their recruitment process or
camp design \cite{Bryant2019AGood, Krug21, Begel21, GRANT2023102773}.
Similarly, the organizing team of this camp incorporated a deliberate
recruitment process towards including students from each parish of Jamaica. Bryant \etal \cite{Bryant2019AGood} describes a camp that emphasizes
data science for social good in order to attract minority groups that
are likely to prioritize careers that improve their communities.
Grant \cite{GRANT2023102773} investigates a coding camp designed
exclusively for girls. Krug \etal \cite{Krug21} discusses a camp
based upon using code to make hip hop beats and appeal to urban
youth of color who are likely to enjoy this genre of music. Begel
\etal \cite{Begel21} aimed to increase Computing exposure for students
with autism by designing and implementing an entirely remote summer camp,
intending to prioritize comfort and accessibility. 

Coding camps have been found to improve systematic problem-solving skills,
interest in coding, and coding-related confidence
\cite{Nugent2016RoboticsProject, GRANT2023102773, Ahmet22}. In particular,
Bryant \etal \cite{Bryant2019AGood} found that self-efficacy increased more
for girls than boys. However, the aforementioned coding camps lasted for only
one week and were not generally designed to introduce algorithms and
algorithmic thinking within their respective curricula
\cite{Nugent2016RoboticsProject, Fields2016CombiningCamp,
Lee2016SummerSchoolers}. Bryant \etal \cite{Bryant2019AGood} did expose
students to algorithms within the context of using data science for social
good, and Chen \etal \cite{Chen19} present a camp that exposed students to
the application of algorithms within the digital humanities. The current
paper similarly diverges from this non-algorithmic trend and discusses a
camp that explicitly sets out to teach high school students college-level
programming and algorithms. This camp is different as it lasts
4 weeks instead of the literature's established standard of 5 days.

Robotics summer camps have been found to boost middle school student interest
in engineering careers \cite{Ahmet22}, possibly because these camps expose youth to material
and skills to which they have not been previously introduced
\cite{Nugent2016RoboticsProject}. By similarly introducing students to
concepts to which they have not yet been exposed, the camp described
in the current
paper sought to increase student interest in related Computing
careers. Research into the previously mentioned robotics program went on to
show that participation in robotics camps, clubs, and competitions improved
robotics related self confidence and promoted STEM learning, including
knowledge of programming. In \Cref{subsec:STEMclubs}, we explore this further
by discussing the impact of prior participation in STEM clubs on programming
and algorithms performance while at camp.

\section{Study Design}
\label{sec:Design}
This section documents our study design. We present an overview of the camp, the camp syllabus,
our research questions and the data collection approach.
\subsection{Camp Overview}
\label{subsec:campContext}

This 4-week summer residential camp focused on teaching high school students
college-level programming and algorithms. The intent was to expose 
students to programming and develop their problem-solving skills.
The academic schedule of camp consisted of lectures followed by lab sections
in which students solved programming exercises related to the concepts introduced
in the lecture. Additional details about camp are provided next.

\paragraph{Camp Staff}
Academic staff consisted of 4 lecturers and 11 teaching assistants (TAs).
Lecturers were senior Computer Science researchers (tenured faculty or
equivalent), with each lecturer
present for exactly one week of camp for
a total of 12 on-site academic staff. Among the TAs, there were seven
graduate-level and four undergraduate-level Computer Science students.
Three academic staff members were local (one lecturer, two TAs), and the
rest traveled from abroad.

Non-academic staff consisted of 8 ``chaperones'' responsible for
the students' safety and well-being. In addition, the camp was supported
by university facilities (administration, cooking, cleaning, and healthcare)
and an organizing committee.

\paragraph{Student Recruitment and Admission}
Students were recruited indirectly by organizers contacting high
schools in all parishes of Jamaica, with the goal of including
students from every parish. School administrators, teachers, and parents
then encouraged students to apply.
Applications were collected in an online portal. Student
applicants submitted transcripts from the last two academic years, a
short essay on why they would like to attend, and
demographic details: name, age, gender, grade, and
high school.
In all, there were 432 applicants and 1 additional student who was
referred to the organizers by a government agency outside of the
application process. Amongst the 432 applicants there were 222 female applicants and
210 male applicants.

Applications were reviewed by TAs and a member of the
organizing team over three rounds. First, applications were split amongst
the TAs and reviewed using a provided rubric evaluating the applicant's
transcript and essay. Based on these, each reviewer
composed a \emph{long list} of potential applicants (the remainder were rejected).
The second round involved each reviewer
evaluating a different long list based on the same rubric,
so that each long-listed application received two reviews.

Reviewers then met to narrow down the cumulative long list,
taking into account each applicant's
parish and high school.
By the end of this process, 50 high school students from across Jamaica were offered admission to the camp (acceptance rate $11.5\%$).
While all 50 students accepted their offer to join the camp, only 48
students actually attended.\footnote{It is unknown why the two students
did not attend.} Given the racial and ethnic makeup of Jamaica's population,
race was not considered during the selection process. However, all
but one of the students were of black origin.

\paragraph{Syllabus}
The camp syllabus was modeled after an affiliated camp, \emph{AddisCoder}, that targets a similar audience
in Ethiopia. A few changes had to be made to accommodate the
different mathematical background of students in the two countries.
\begin{enumerate}
 \item Python programming constructs (types, operators, conditional statements, loops, functions, lists and dictionaries).
 \item Recursion.
 \item Big $O$ notation and asymptotic running time analysis.
 \item Searching and sorting algorithms (linear search, binary search, selection sort, and merge sort).
 \item Graphs (breadth-first search and depth-first search).
\end{enumerate}

See \url{https://jamcoders.org.jm/syllabus/2023/} for a full syllabus, lecture notes, exercises and exams.

\paragraph{Lectures and Labs}
The academic portion of camp took 8 hours/day, including breaks and
transitions, 
and consisted of two sessions (morning and afternoon), with 
an hour-long lecture and two hours of lab in each.
Each week, lectures were given by a different lecturer.

Lab sessions were led by the TA team. Each lab would start with a 
recap session reiterating concepts needed for the lab, and an
open discussion about any concepts the students struggled with
in the preceding lecture.
Following this, students would complete exercises on 
individual computers, during which they could
ask for help from the 11 TAs by raising their hand.
If multiple students 
seemed to be interested in hearing clarifications about certain topics 
while attempting the lab exercises, the TAs would offer impromptu sessions 
in a separate classroom. This meant that 
students could elect to be exposed to the material multiple times,
in a variety of settings, in order 
to cement their understanding.

\paragraph{Quizzes and Study Halls}
Students received three quizzes during the camp. The first quiz, which
we called quiz 0, was an ungraded mock exam held in Week 2. The purpose of this quiz
was to prepare students for quizzes 1 and 2, which were graded. These
quizzes were administered in weeks 3 and 4.
In the evening before each quiz, optional ``Study Hall'' sessions were held outside of the student dorms.
In these open-ended revision sessions, students could ask the TAs present questions about any portion
of the material (as opposed to labs, which were focused on the preceding lecture).
The five commuting students were accommodated via a
virtual option facilitated by one TA.

\paragraph{Big Siblings}
At the end of Week 1, each student was assigned a TA as their ``Big Sibling.''
TAs would then meet their ``Little Siblings'' (in groups or individually) in
informal sessions meant to complement the technical guidance given in lab.
The goal was to make sure that each student has at least one TA they are
comfortable talking to about \emph{any} emotional and academic challenges
that arise in this academically-intensive camp. On the other hand, the
small number of Little Siblings allocated to each TA (4--5 per TA)
ensured that the academic team was generally aware of the state of each
student and could provide personalized support.

\paragraph{Academic Staff Meetings}
At the end of each academic day, TAs and the on-site lecturer met
for a \emph{debrief} to reflect on the day's events and make adjustments for future days.
Additionally, each week was preceded by an academic all-hands meeting of TAs with the incoming
lecturer to report on the camp status and finalize content for the upcoming week.

\subsection{Research Questions}
\label{subsec:rq}
The following research questions guided our study of this 4-week algorithms and coding camp for high school students:
\begin{description}
 \item [RQ1.] Are there gender differences in performance between
 male and female students?
 \item [RQ2.] Does school location or school ranking
 have an impact on student performance?
 \item [RQ3.] How does participation in a STEM club at a school affect
 student performance? 
 \item[RQ4.] What are the changes in students' attitudes toward
 studying Computing after taking part in the camp?
\end{description}

\subsection{Data collection}
During the camp the students received two graded quizzes (at the end
of Weeks 3 and 4). The quizzes were taken by all attending students:
48 for the first quiz and 47 for the second.%
\footnote{One student could not attend the second quiz due to a severe illness.}
Each quiz covered the entire material taught until its date, and was scored
out of 100 points.

Our dataset consists of quiz scores aligned with student demographic information.
Quiz scores are our only ``traditional'' metric for performance in the camp.

Following the camp, a survey was sent to students collecting feedback about camp and additional demographic information.
Of the 48 students, 24 filled out the survey and consented to analysis of their responses.%
\footnote{One respondent completed the survey but did not consent to having their data analysed; it is excluded from our dataset.}

\section{Results}
\label{sec:rslts}
In this section we present the results showing student performance on the
quizzes administered during the camp. These results are presented to
highlight the answers to the research questions posed in
\Cref{subsec:rq}. We rely on nonparametric statistical tests for
our analysis given that the quiz scores are not normally distributed.

\subsection{Gender Differences}
\label{subsec:SexDiffs}
\begin{figure}[!h]
\centering
\includegraphics[width=0.95\linewidth]{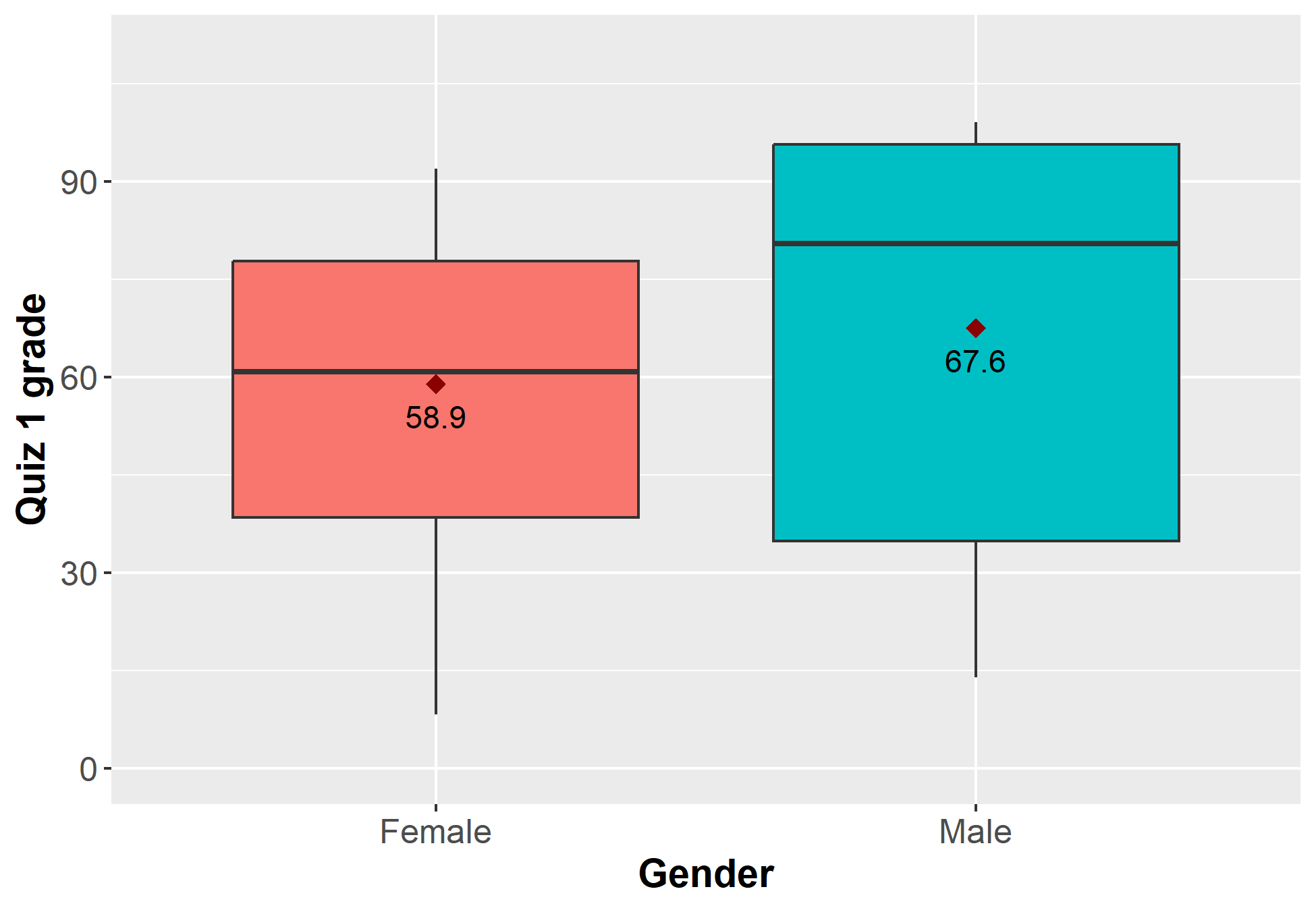}
\caption{Boxplot showing quiz 1 scores by gender}
\label{fig:bpQuiz1Gender}
\end{figure}

Of the 48 students who took part in the camp, 28 were female, and 20 were
male. To answer RQ1, we examined two boxplots of Quiz~1 scores grouped
by gender, as shown in \Cref{fig:bpQuiz1Gender}. The numbers within
each boxplot represent the mean for each group.

From \Cref{fig:bpQuiz1Gender}, it is clear that the median score on Quiz~1 for
male students was greater than that for female students. The same was true for the
variations in scores (larger variation among male students). 
However, a Wilcoxon signed rank test showed that the difference in median scores
between the genders was not statistically significant
($W = 351.5; Z=1.5; p = 0.1375$). A similar analysis for Quiz~2 yielded the
same comparisons.

\begin{figure}[!t]
\centering
\includegraphics[width=0.95\linewidth]{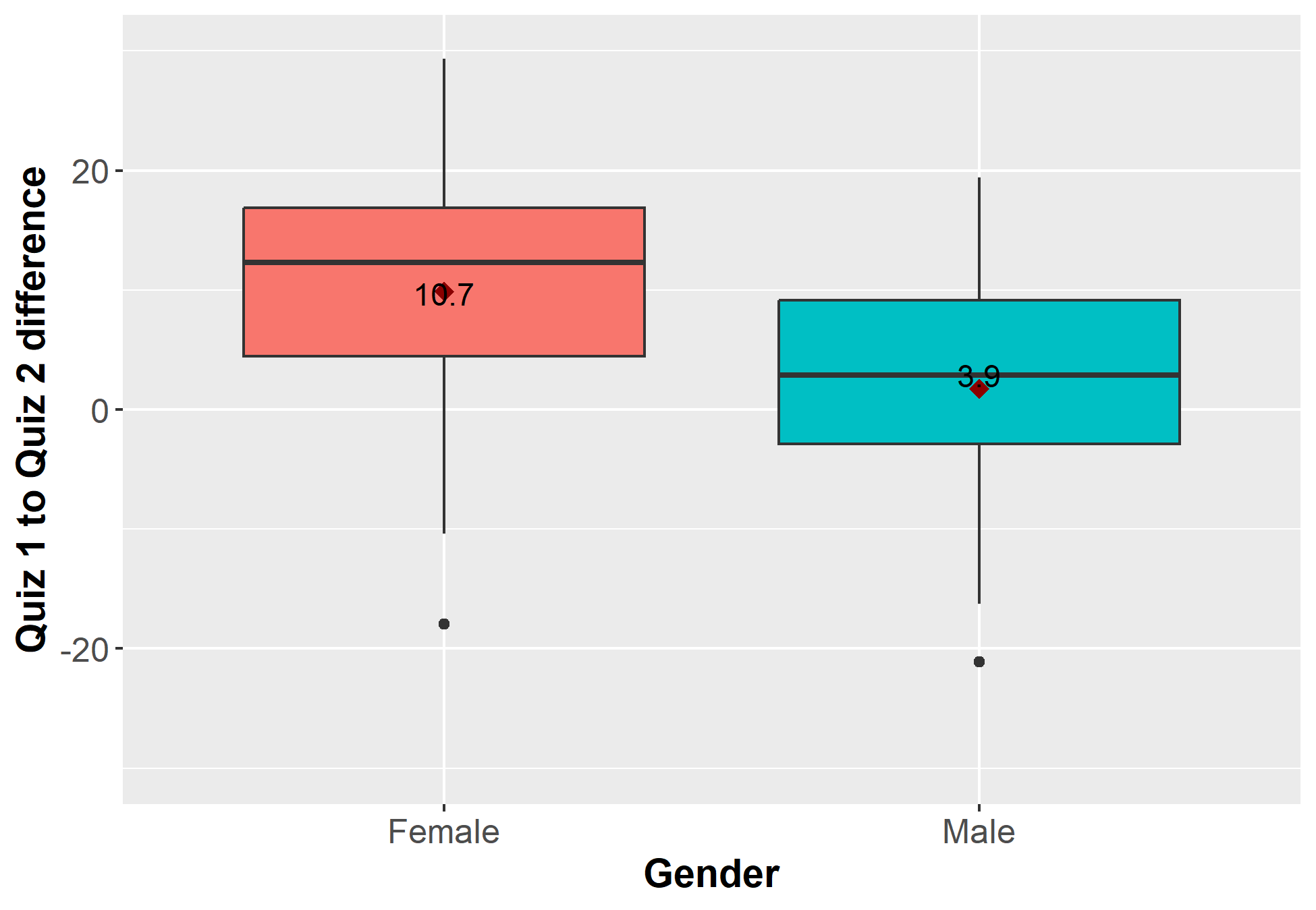}
\caption{Boxplot showing improvement in scores from Quiz~1 to Quiz~2 by gender}
\label{fig:bpQuiz1toQuiz2Sex}
\end{figure}

\Cref{fig:bpQuiz1toQuiz2Sex} shows per-student performance improvement. While
the performance improvement for female scores \emph{increased} by about 10.7
points, male student scores \emph{decreased} by about 1.5 points.
We provide one possible explanation for this noticeable difference
in \Cref{sec:Discuss}.

A Wilcoxon signed rank test showed that there was a
statistically significant difference ($W = 167; Z=-2.36; p = 0.0176$)
between the median improvement in scores between Quiz~1 and Quiz~2 for female
and male students.

\subsection{Impact of School Location and Rank}
\label{subsec:schoolImpact}

\begin{figure}[!t]
\centering
\includegraphics[width=0.95\linewidth]{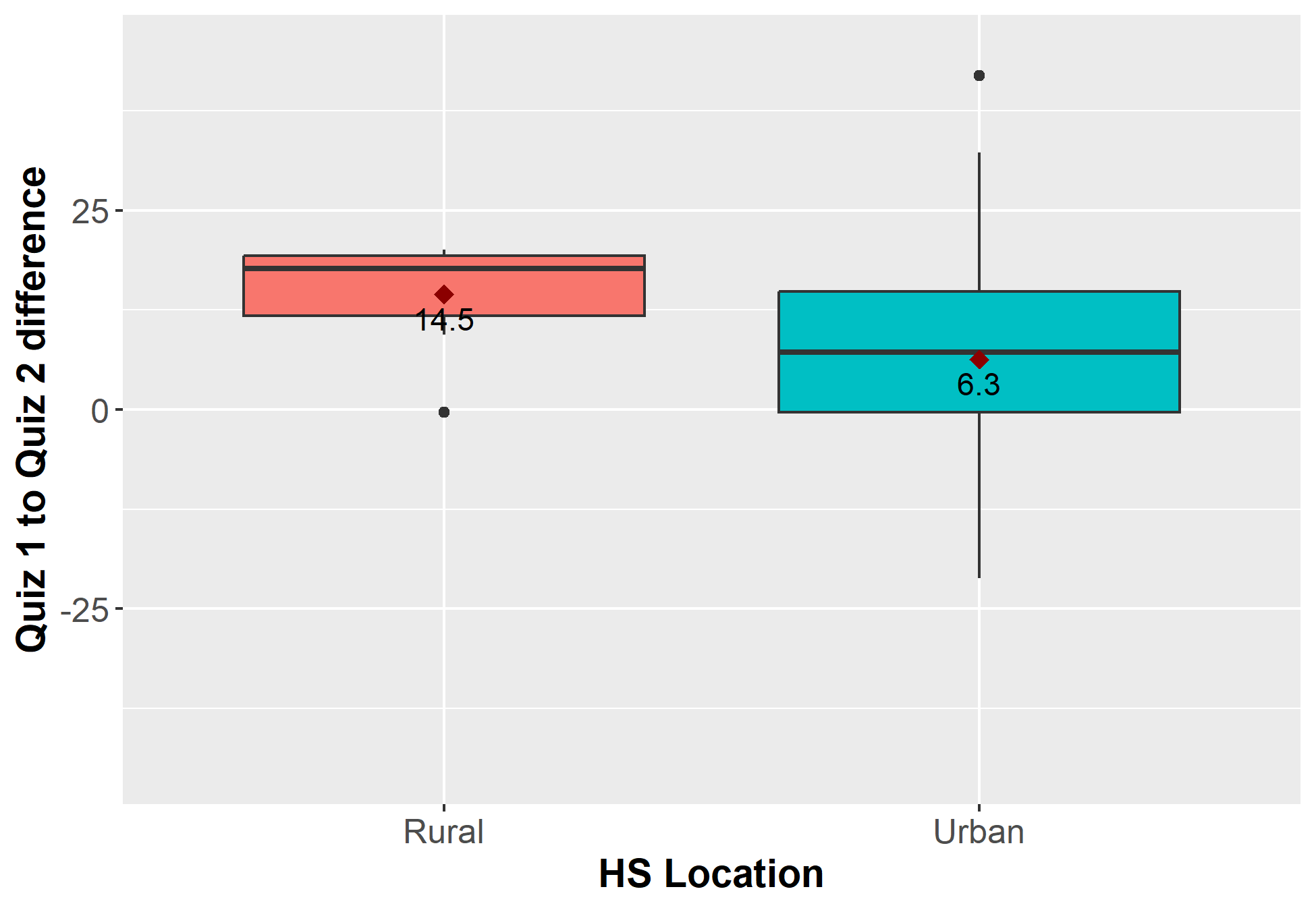}
\caption{Boxplot showing performance improvement from Quiz~1 to Quiz~2 by high school location}
\label{fig:bpQuiz1toQuiz2HSLoc}
\end{figure}

\paragraph{School Location}
The Jamaican Ministry of Education classifies high schools as being in either
rural or urban locales. Of the 48 students selected, 40 attended an urban
high school, 7 a rural school, and 1 attended a foreign school in another
small island developing state. This foreign student was
excluded from the analysis in this section.
The median Quiz 1 grade for urban students was 74.7 whereas the 
corresponding statistic for rural students was 37.8. A Wilcoxon signed
rank test showed that there was a
statistically significant difference ($W = 219; Z=2.36; p = 0.0190$)
between the median score for urban and rural students.
In Quiz 2, urban students outperformed rural students with a statistically significant difference
($W=211; Z=2.12; p=0.0351$).
\Cref{fig:bpQuiz1toQuiz2HSLoc} shows the performance improvement
in scores from Quiz~1 to Quiz~2. \Cref{fig:bpQuiz1toQuiz2HSLoc} shows
that the average performance improvement in scores from Quiz~1 to Quiz~2
was about +14.5 points for rural students, whereas on average the urban
students scored about +6.3 points more on Quiz~2 compared to Quiz~1. 
Furthermore, this performance improvement was statistically significant
($W = 70; Z=-2.38; p = 0.0160$). As a
result, we conclude that the rural students improved more between the quizzes.
In \Cref{sec:Discuss}, we will provide one argument for this 
observation. We also examined the differences between students who
participated in STEM clubs in schools versus those who did not. Due to 
constraints of a small data set, we did not find statistically significant differences. 

A Wilcoxon signed rank test showed that there was a
statistically significant difference ($W = 69; Z=-2.12; p = 0.0330$)
between the median performance improvement from Quiz~1 to Quiz~2 for urban and rural
students.

\begin{figure}[!t]
\centering
\includegraphics[width=0.95\linewidth]{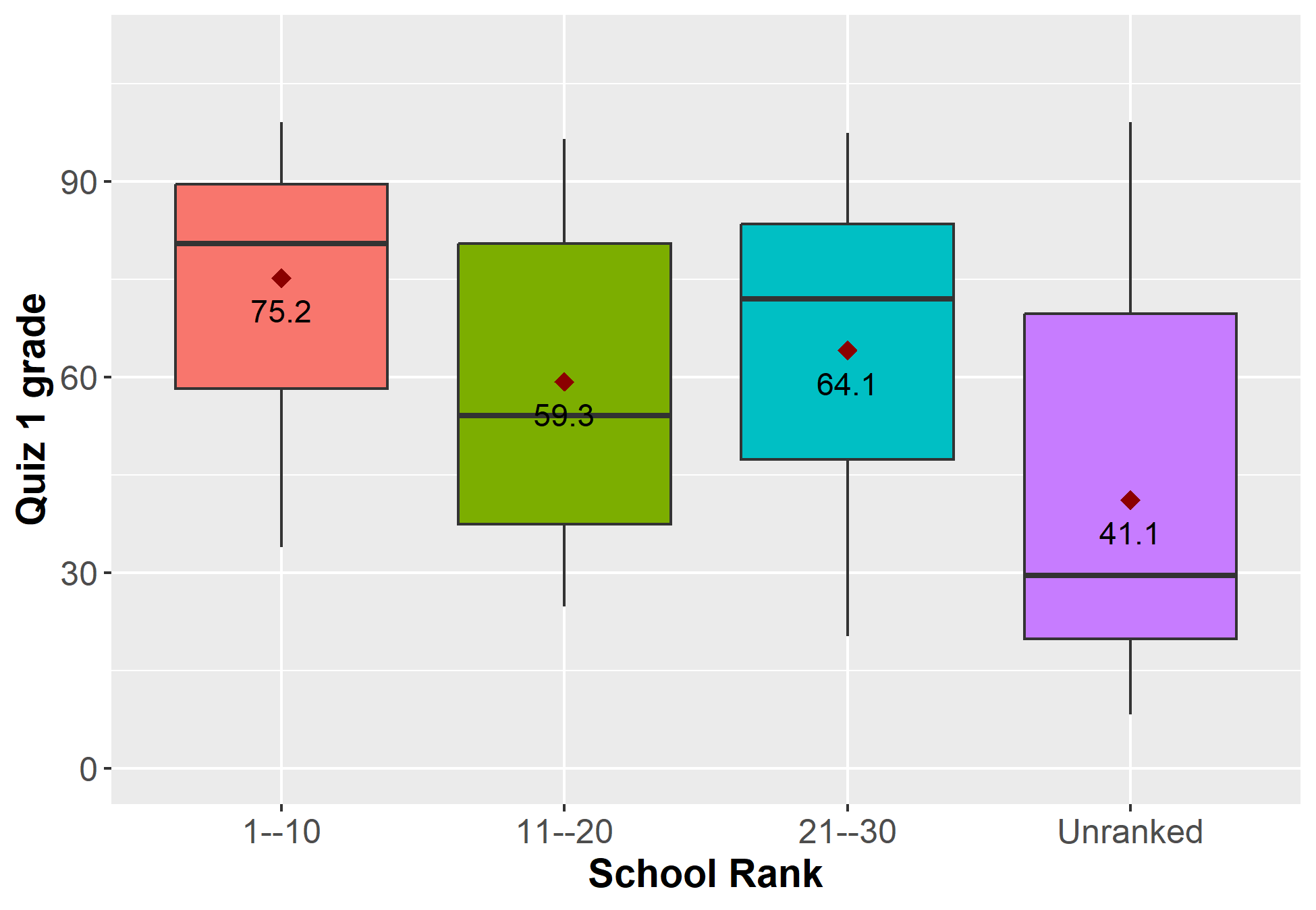}
\caption{Boxplot showing Quiz 1 scores by school rank}
\label{fig:bpQuiz1SchoolRank}
\end{figure}

\paragraph{School Rank}
Within Jamaica, a nongovernmental organization has
ranked high schools at periodic intervals based on student performance in
regional examinations. Some schools are excluded from the rankings if
the school does not meet the NGO's minimum requirement for
ranking.%
\footnote{The nongovernmental organization ranks high schools on
the basis of the institution's cohort leaving grade 11 with passing scores
in  five (or more) subjects, including mathematics and English, at one
sitting. Schools that have less
than 50\% of the cohort leaving grade 11 with five or more subjects,
including mathematics and English, are excluded from the ranking.}
For the camp, 22 students came
from schools ranked between 1 and 10, 8 from schools ranked between 11 and
20, 5 from schools ranked between 21 and 30, and 12 from unranked
schools. \Cref{fig:bpQuiz1SchoolRank} shows boxplots
with Quiz~1 grades and the school rank.

From \Cref{fig:bpQuiz1SchoolRank}, it is seen that the median score for
students declines as the school rank decreases. The exception is
schools that are ranked between 11 and 20. Repeated pairwise Wilcoxon signed
rank tests show that a statistically significant difference in the median
Quiz~1 scores only existed between those students from the top-ranked and
unranked schools ($W = 219.5; Z=3.15; p = 0.0017$).

\subsection{Impact of a STEM Club}
\label{subsec:STEMclubs}
High schools in Jamaica have different STEM-related clubs for students, e.g.,
math, Computing/software engineering, robotics, engineering, environmental,
chemistry, and medical study based clubs. To answer RQ3, we examined
student report cards
and application essays for evidence of participation in these clubs. The review of student records
showed that 18 students participate in STEM clubs, while 30 did not.

\begin{figure}[!t]
\centering
\includegraphics[width=0.95\linewidth]{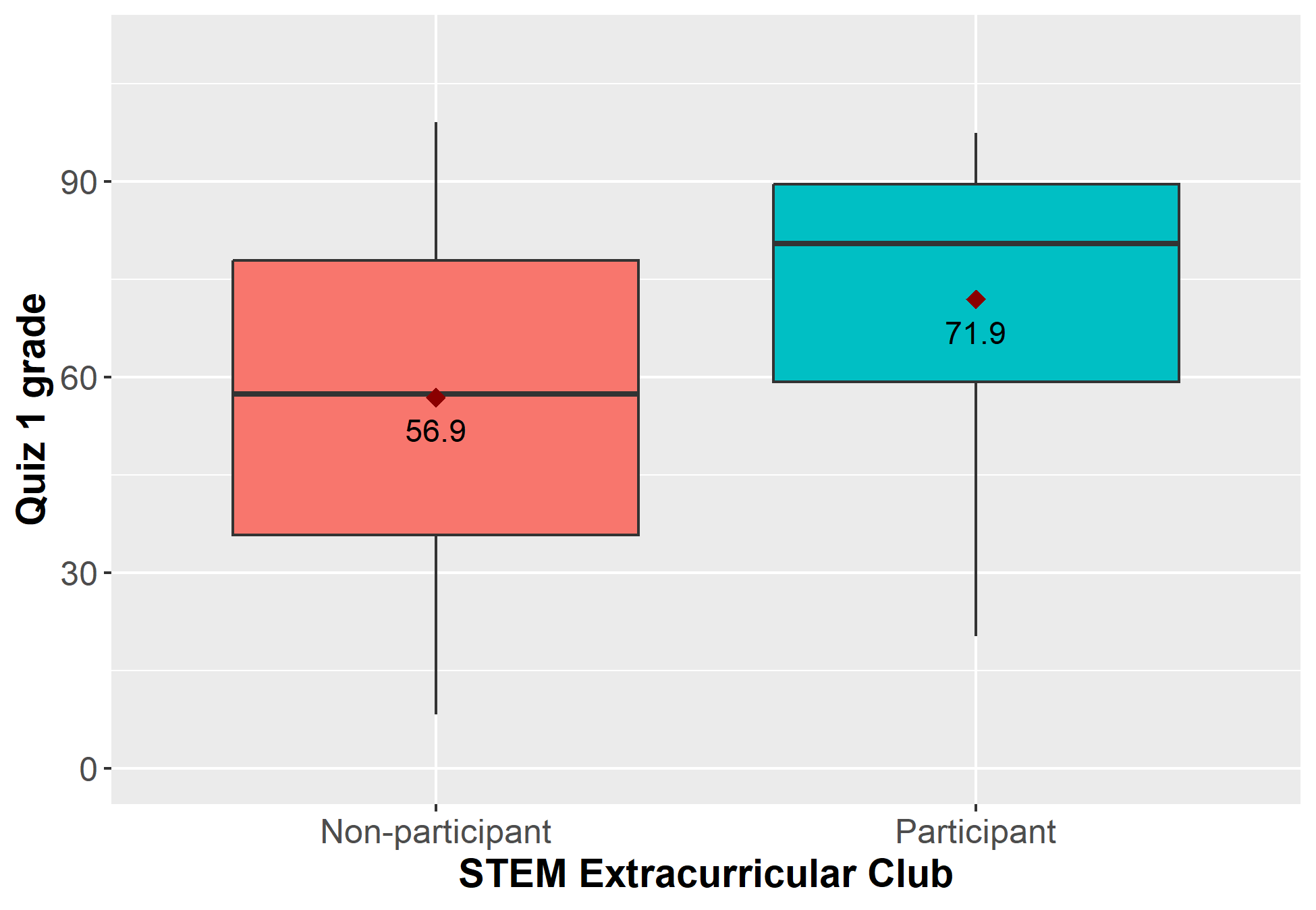}
\caption{Boxplot showing Quiz 1 scores by STEM club participation}
\label{fig:Quiz1bpSTEMClub2023}
\end{figure}

From \Cref{fig:Quiz1bpSTEMClub2023} it is clear that students who
participated in a STEM club at school generally performed better on Quiz~1 than
non-participants. The median Quiz~1 grade for
students participating in STEM clubs was $80.4$\% whereas the corresponding
statistic for non-participants was $57.4$\%. A Wilcoxon signed
rank test showed that there was not a statistically significant difference
($W = 353; Z=1.77; p = 0.0789$) between the median score for participants and
non-participants in STEM clubs. Similar analysis for Quiz 2 also
reflected no statistically significant difference ($W=361; Z= 1.94;
p = 0.0539$).

The average performance improvement from Quiz~1 to Quiz~2 was about +5.6
(median 11.4) points for students who were non-participants in STEM clubs,
whereas on average the STEM club participants scored about +5.9 points
(median 4.6) more on Quiz~2 compared to Quiz~1. A Wilcoxon signed rank test
showed that there was not a statistically significant difference
($W = 213; Z=-1.21; p = 0.2314$) between the median performance improvement
from Quiz~1 to Quiz~2 for STEM club participants and non-participants.

\subsection{Changes in Attitude toward Studying Computing}
To answer RQ4, we conducted a post-camp survey (24 responses) which included these two questions: (a) state your 
preferred majors in college/university before the camp, and (b)
state your desired major in college/university after the camp. Nine
students responded that they wanted to major in Computer Science or a
Computing-related discipline prior to the camp. Following the camp, 14
students said that they wanted to major in Computer Science or a
Computing-related discipline. Looking at the results in greater detail, it
was seen that six students who did not want to major in Computing prior
to the camp thought that they would like to major in Computing after the
camp. However, there was one student who wanted to major in Computing prior
to the camp, who then indicated that he/she wanted to study a
non-Computing discipline after the camp. These two numbers amount to
a net increase of five students who wanted to major in Computing disciplines in
university/college following the camp. We conclude that the
camp had a positive impact on student interest in pursuing higher education in
Computing.

\section{Discussion}
\label{sec:Discuss}

Next, we expand on the quantitative analysis of \Cref{sec:rslts} with
additional perspectives from the camp staff.

\paragraph{Big Siblings as urban-rural equalizers}
The Big Sibling program was determined to be a success, especially for students from schools ranked 21--30 or unranked. A significant amount of time and effort, during and outside the classroom, was invested in students who were slower to understand early topics. Indeed, TAs assigned as Big Siblings to these students directed more of their efforts towards tutoring these students on early topics, averting an unbridgeable gap later on.

This qualitative testimony is supported by our analysis in \Cref{subsec:schoolImpact}: students from rural areas were slower to understand early topics (lower Quiz 1 score) and, therefore, received increased tutoring from their Big Siblings. The result: students from rural areas improved 14.1 points between quizzes, compared to just a 3.9-point improvement for students in urban areas.

\paragraph{Why did female students improve more than male students?}
Female students improved significantly more than their male counterparts between quizzes (+10.7 versus -1.5 points, respectively). Camp TAs could not provide a qualitative explanation,%
\footnote{In particular, TAs reported that female students did \emph{not} receive noticeably more tutoring than male students, unlike the urban-rural case above.}
and so we turn to the literature:
Female student population in other Caribbean islands are underrepresented in higher education Computing programs \cite{FokumCL16}, yet are known to academically outperform their male counterparts at a general high-school level \cite{thompson2017planning}. Could it be that the same societal factors that inhibit female student success in Computing positioned them at a disadvantage at the start of our camp, yet their academic ability helped them bridge the gap by the end of it? At the very least, we conclude that
camp was a comfortable environment for female students to learn and improve.%
\footnote{To quote a female student: ``This camp not only teach me about coding and algorithm, but is a \emph{safe space} for children my age that have a common interest in technology.''}

\paragraph{Changes in attitudes towards studying Computing}
It is worth noting that the camp resulted in a net gain of five students who wanted to
major in Computer Science in university/college. Two of the five students come from
schools where they are not required to take ``Information Technology'' for the grade 11
regional examinations. In addition, their schools do not offer Computer Science at the
grade 13 regional examinations. Thus, the coding camp provides students to high quality
Computer Science instruction, and it also attracts more students to the discipline.

\paragraph{Heterogeneity affords gaps}
The camp organizers explicitly sought to recruit a diverse pool of students in terms of gender, age, parish, school locale (urban/rural) and rank. As seen in \Cref{sec:rslts}, these demographics were sometimes noticeably correlated with quiz score in camp. Throughout camp, the academic staff emphasized repeatedly to the students that quiz scores should not be viewed as the main outcome of camp, but rather as a form of feedback (indeed, these scores are not reported back to the high schools or parents). We ask our reader to do the same: scores are convenient for statistical analysis, but students that scored below average did not, by any means, ``fail'' at camp.\footnote{Consider a student that initially refused to take the first quiz for fear of earning a low score; a lengthy conversation with her Big Sibling conveyed that attempting a significant challenge and failing is, sometimes, more rewarding than not facing it at all. The student took the first quiz and indeed earned the lowest score in class. But in the second quiz, the student improved threefold! On graduation day, neither staff nor the student viewed the student's performance in camp as a failure.}

\section{Limitations}
\label{sec:limits}
\paragraph{Data collection} The \emph{STEM club participation}
demographic may be underreported in our dataset, as we manually
collected it from essays and transcripts but not all transcripts
explicitly report this.
Our reliance on two quiz grades as a sole metric is fundamentally 
limited by the imperfection of examination as a measurement of 
understanding (see, e.g., an early concern in \cite{OnExams}).

\paragraph{Author positionality} This paper was authored by some
members of the academic staff or organizing team of the camp. All but
one author are Jamaican locals and affiliated with the university in 
which the camp was located. This deep and personal understanding of
the camp and its context may be viewed as either an advantage or a
disadvantage (or both).

\section{Conclusions}
\label{sec:concl}
Our work examines a summer camp that teaches college-level algorithms and programming to high school students in grades 9--11. The camp successfully included demographically diverse students, with 15\%/83\% students from rural/urban schools,\footnote{There was a single student from a foreign school.} and 58\%/42\% female/male students, respectively. We found that female students improved significantly more over the duration of the camp than their male counterparts, and likewise for rural versus urban students. This demonstrates that camp can be an effective and empowering exposure to Computing, even in a heterogenous environment.

A low student-teacher ratio can help maximize student success \cite{KOC201565,Meads0D23}. We found this to be particularly true when examining the impact of the Big Sibling program on the rural-urban divide in student performance.
A low student-teacher ratio allowed each Big Sibling to give personalized tutoring to all of their assigned students, with no student left behind.

We conclude that introducing advanced Computer Science concepts at a young age is possible, even in a student population of mixed background and life opportunities. A low student-teacher ratio was instrumental in identifying and supporting at-risk and underrepresented students. We recommend the inclusion of a similar Big Sibling program in any coding camp with sufficient academic staff, especially when the camp population is demographically heterogeneous.

A main goal of this camp was to introduce Computing as a viable career option (or at the very least, area of interest) to students in an island whose population is almost-entirely underrepresented in the field of Computer Science.
Our post-camp survey found that 58\% of respondents would like to major in Computing at a tertiary level. This includes 25\% of respondents who previously did not want to major in Computing.%
\footnote{Rest assured, only one respondent who previously wanted to major in Computer Science changed their mind after camp.}

This profound change imparts the impact of such a camp, which offers students from all backgrounds the opportunity to gain Computing exposure for free and with the academic support required to bridge existing knowledge gaps. If well-delivered, these camps can result in even more people from diverse backgrounds being drawn into the discipline.

\section*{Acknowledgements}
We thank JamCoders donors, staff and students for making this camp possible. %
In particular we thank Jelani Nelson for helpful comments on an early version of this report. %

\bibliography{references}

\newpage
\appendix
\section{Figures}

\newcommand{\categorical}{Categorical}
\newcommand{\tablespace}{\rule{0pt}{4ex}}
\begin{figure}[h!]
  \label{tab:varList}
  \centering
  
 \begin{tabular}{c c c c c}
    \toprule
    \textbf{Variable} & \textbf{Data type} & \textbf{Data values} & \textbf{Mean} & \textbf{Median} \\
    \midrule
    Quiz 1 score & Float & 8.3--99.1 & 62.5 & 70.7 \\
    Quiz 2 score & Float & 15.8--98.3 & 69.7 & 73.3\\
    Performance improvement & Float & -21.1--41.8 & 7.9 & 8.4 \\
    \tablespace & & & \multicolumn{2}{c}{\textbf{Mode (count)}} \\
    \cmidrule(lr){4-5}
    Gender & Categorical & Male, Female & \multicolumn{2}{c}{Female (28)}\\
    Grade & Categorical & 9, 10, 11 & \multicolumn{2}{c}{11 (26)}\\
    Residence type & Categorical & Commuting, Dorm & \multicolumn{2}{c}{Dorm (30)}\\
    School locale & Categorical & Rural, Urban & \multicolumn{2}{c}{Urban (40)}\\
    School rank & Categorical & 1--10, 11--20, 21--30, Unranked & \multicolumn{2}{c}{1--10 (22)}\\
    STEM club participant & Categorical & Participant, Non-participant & \multicolumn{2}{c}{Non-participant (30)}\\
    \bottomrule
\end{tabular}
\caption{List of variables and summary statistics. Mean and median is reported for float type variables, and mode reported for categorical variables. Performance improvement is defined as the score on Quiz 2 minus the score on Quiz 1.}
\end{figure}

\newcommand{\figwidth}{0.75\linewidth} %

\end{document}